\documentclass[pre,preprintnumbers,amsmath,amssymb,twocolumn, nofootinbib]{revtex4}
\usepackage{graphicx}
\usepackage{bm,color}
\usepackage{mathrsfs}
\usepackage{verbatim}

\definecolor{labelkey}{cmyk}{.4,.2,0,0}

\renewcommand{\log}{\ln}

\begin{document}

\title{Role of the glassy dynamics and thermal mixing in the dynamic
nuclear polarization and relaxation mechanisms of pyruvic acid}
\author{M. Filibian$^{1}$, S. Colombo Serra$^{2}$, M. Moscardini$^{1}$, A. Rosso$^{3}$, F. Tedoldi$^{2}$ and P. Carretta$^{1}$}

\affiliation{\medskip
$^{1}$ University of Pavia, Department of Physics, Via Bassi 6, 27100-Pavia, Italy\\
$^{2}$Centro Ricerche Bracco, Bracco Imaging Spa, via Ribes 5, 10010 Colleretto Giacosa (TO), Italy. \\
$^{3}$Universit\'e Paris-Sud, CNRS, LPTMS, UMR 8626, Orsay F-91405, France.\smallskip}
\begin{abstract}
The temperature dependence of $^1$H and $^{13}$C nuclear
spin-lattice relaxation rate $1/T_1$ has been studied in the  1.6
K - 4.2 K temperature range in pure pyruvic acid and in pyruvic acid containing
trityl radicals at a concentration of 15 mM. The temperature
dependence of $1/T_1$ is found to follow a quadratic power law for
both nuclei in the two samples. Remarkably the same
temperature dependence is displayed also by the electron spin-lattice
relaxation rate $1/T_{1e}$ in the sample containing radicals. These results
are explained by considering the effect of the structural
dynamics on the relaxation rates in pyruvic acid. Dynamic nuclear polarization
experiments show that below 4 K the $^{13}$C build up rate scales
with $1/T_{\text{1e}}$, in analogy to $^{13}$C $1/T_1$ and consistently
with a thermal mixing scenario where all the electrons are collectively involved in the dynamic nuclear polarization process and the nuclear spin reservoir is in good thermal contact with the electron spin system.
\end{abstract}

\maketitle
\section{Introduction}
In recent years Dynamic Nuclear Polarization (DNP) has been shown to be one
of the most promising technique for hyperpolarizing nuclear
spins. DNP increases the nuclear steady state
polarization thanks to a transfer of magnetic order from the electron
to the nuclear spins under microwave irradiation close to the
electron Larmor frequency ($\omega_e$). The application of DNP has
catalyzed major advances in the Nuclear Magnetic Resonance (NMR) of low sensitivity nuclei in nanosized materials \cite{emsley1}, in the high resolution NMR of biological samples \cite{raey, griffin2} and in {\em in vivo} real time imaging of
biomolecules, hardly achievable
with other methods \cite{DUTTA}. For preclinical Magnetic
Resonance Imaging (MRI) DNP is performed in solutions containing
diamagnetic biomolecules labelled with $^{13}$C and a small
concentration of stable radicals. The mixture is cooled down to
about $1$ K and, once the maximum $^{13}$C polarization is
reached, it is rapidly dissolved \cite{Ardenkjaer-Larsen2003,
Wolber2004173, Comment} and injected {\em in vivo}, where the
metabolic processes accessed by the hyperpolarized substrates are
monitored by means of $^{13}$C MRI or Spectroscopy
\cite{DUTTA,Golman25072006,Golman15112006}.

While significant scientific and technological efforts are
nowadays spent to introduce dissolution DNP into the clinical
practice \cite{Kurhanewicz,lin2014,cardiac2013}, there is
growing interest in the fundamental investigation of the physical mechanisms
driving DNP. The first basic description of the DNP phenomenology
dates to a few decades ago \cite{abragorder}, when different
regimes, the Solid Effect, the Cross Effect and the Thermal Mixing
(TM), were defined depending on the magnitude of parameters such as the nuclear resonance
frequency ($\omega_L$), the coupling among the electron and nuclear spins and
the external magnetic field strength. The most common and relevant
regime for the molecules utilized in metabolic imaging is
seemingly the TM \cite{Ardenkjaer-Larsen2003, Comment,ganiso}, which is effective when the
electron resonance linewidth is larger than $\omega_L$ and the interactions among nuclear and electron
spins are large enough to establish a common spin temperature
for the two reservoirs.

The TM regime is attained in pyruvic acid (PA) labelled with
$^{13}$C and doped with a concentration of trityl radicals  ($c$)
of the order of 10 mM \cite{ganiso, Johannesson2009,macholl2010}.
PA has been up to date the most widely investigated system for
{\em in vivo} DNP applications due to its role in glycolytic
pathways \cite{Golman25072006, Golman15112006}
and it can be considered as a prototype system to study the TM
regime. Several DNP experiments have shown $^{13}$C
polarizations approaching 20-30 \%  in PA doped with trityl
radicals, at a temperature $T\simeq 1.2$ K and for a magnetic
field ($H$) of 3.35 Tesla
\cite{Ardenkjaer-Larsen2003,Golman25072006}. In order both to
optimize and to validate novel theoretical models of TM, several
investigations of the nuclear and electron relaxation processes
around 1.2 K have been performed. The effect of relevant
parameters, including the radical concentration  \cite{ganiso,
Johannesson2009, macholl2010}, the concentration of gadolinium
complexes \cite{ganiso, Lumata2012}, the nuclear
concentration \cite{Lumata2011,colombo3}, the amount of matrix
deuteration \cite{Lumatadeut}, the effect of microwave saturation
and the field strength \cite{Johannesson2009, macholl2010,
Meyer2011} on the DNP performances of this molecule have been
experimentally studied. Remarkably very recently, the
relevance of these physical quantities on DNP kinetics has also
been considered in the development of novel models describing TM
throughout a rate equation approach
\cite{colombo1,colombo2,colombo3}. Nevertheless, the role of the properties of the glassy matrix formed by
the polarized molecules and radicals has not been investigated to a deep level.

The importance to achieve a glassy matrix, yielding a homogeneous
distribution of internuclear and electron-nuclear distances in
order to optimize DNP, has been well recognized \cite{Lumata2011,raey,griffin3}
but a detailed study of the glassy dynamics of PA below 5 K
and its effect on DNP has not been addressed up to date. In this
regard the investigation of the spin dynamics of nuclei
such as $^1$H, not involved in TM, can eventually help to identify
the relaxation processes driven by the coupling with the glassy
dynamics.

In this paper a $^1$H and $^{13}$C NMR
study of pure PA and PA containing trityl radicals at a concentration
of 15 mM is presented. It is shown that the spin-lattice relaxation rate ($1/T_{\text{1}}$) of $^1$H and $^{13}$C
nuclei and of the radical electron spins show all a nearly
quadratic $T$-dependence below 4.2 K. Remarkably, while the $^1$H spin-lattice relaxation rate ($1/T_{\text{1H}}$) is
scarcely affected by the presence of paramagnetic radicals,
the $^{13}$C spin-lattice relaxation rate ($1/T_{\text{1C}}$) shows a sizeable enhancement upon paramagnetic doping. Moreover, the $^{13}$C
polarization build up rate is found to follow the same $T$
dependence of the spin-lattice relaxation rates. All these results are explained, below 4 K, in
terms of the glassy dynamics which characterizes the PA and by
resorting, for $^{13}$C DNP and spin-lattice relaxation, to the TM approach in the regime of good thermal
contact between the nuclear and electron spin systems.

\section{Experimental methods and technical aspects}


1-$^{13}$C pyruvic acid (PA) and un-labeled pyruvic acid (uPA) were purchased by Sigma Aldrich. The free radical trityl OX063 (tris{8-carboxyl-2,2,6,6-benzo(1,2-d:5-d)-bis(1,3)dithiole-4-yl}methyl sodium salt) was kindly provided by Albeda Research. For NMR and DNP experiments, 100 $\mu$L of PA and of uPA, a 15 mM solution of OX063 in 100 $\mu$L of 1-$^{13}$C pyruvic acid (PA15) and a 15 mM solution of OX063 in 100 $\mu$L of unlabelled pyruvic acid (uPA15) were transferred inside quartz tubes and sonicated for 10 minutes. The samples were cooled down to 4.2 K following several procedures, detailed in Appendix 6.1.

DNP experiments were performed by means of a homemade polarizer. A
DNP-NMR probe was inserted in a bath cryostat and placed inside a
superconducting magnet. Within that apparatus the temperature
could be carefully controlled through helium adiabatic pumping
between 1.6 K and 4.2 K. DNP was achieved by irradiating the
samples with microwaves (MW) emitted by a Gunn-diode source
operating in the 96-98 GHz frequency range, with a nominal output
power of 30 mW. $^{1}$H and $^{13}$C NMR probe radiofrequency (RF)
circuits were tuned at 37.02 MHz and accordingly the static magnetic field $H$ was set to 0.87
Tesla and to 3.46 Tesla, respectively. The NMR signals were
acquired with a solid-state Apollo Tecmag NMR spectrometer coupled
to a homemade RF probe.
\begin{figure}[h!]
\centering
  \includegraphics[height=6.8cm]{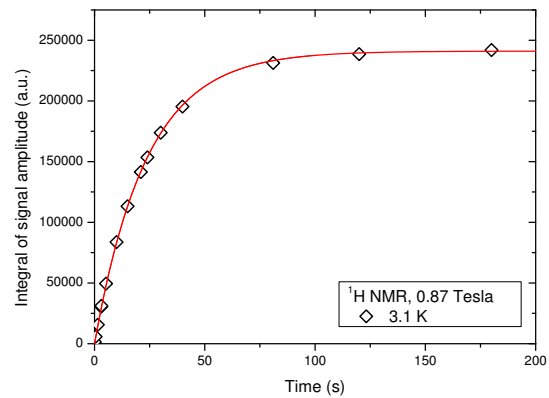}
  \caption{Recovery law for $^{1}$H nuclear magnetization in PA at 3.1 K and 0.87 Tesla after a
  saturating pulse sequence. The solid red line is the best fit according to the function $y(\tau)=M_0(1-\exp(-\tau /T_{\text{1H}}))$.}
  \label{fgr:t1_recovery}
\end{figure}
\begin{figure}[h!]
\centering
\includegraphics[height=7cm]{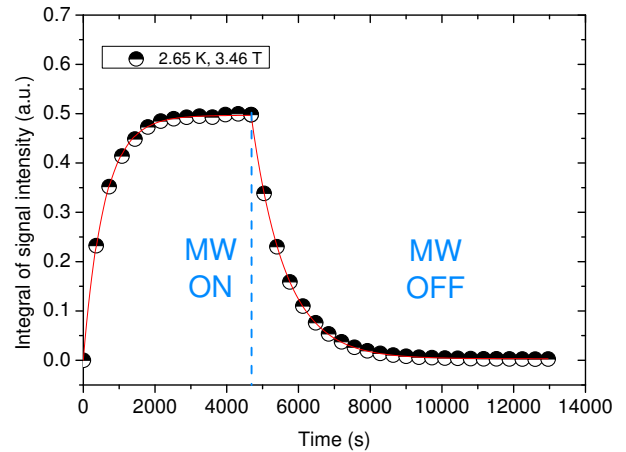}
  \caption{$^{13}$C polarization build-up under MW irradiation (MW ON) and relaxation to the
  thermal equilibrium value of the nuclear magnetization (MW OFF) in PA15 at 2.67 K and 3.46 T. The red lines are fits according to the functions explained in the text.}
  \label{fgr:buildup}
\end{figure}
$^{1}$H NMR spin-lattice relaxation time ($T_{\text{1H}}$) was
measured using standard saturation recovery sequences with a $\pi
/2$ pulse length in the $2-3$ $\mu$s range. In all samples $^{1}$H
recovery law was described by $y(\tau)=M_0[1-\exp(-\tau
/T_{\text{1H}})]$ (Fig. \ref{fgr:t1_recovery}), indicating that
all the protons are characterized by a common spin temperature. $^{1}$H and
$^{13}$C  NMR spin-spin relaxation times ($T_{\text{2H}}$ and $T_{\text{2C}}$)
were measured by means of the Hahn Echo sequence.

DNP experiments were performed by irradiating the sample at the MW
frequency maximizing the positive polarization enhancement, about
97 GHz at 3.46 Tesla.  In order to acquire $^{13}$C build up curves,
the $^{13}$C NMR signal was sampled under MW irradiation after RF
saturation (Fig. \ref{fgr:buildup}). The Free Induction Decay
(FID) signal was acquired up to steady state applying subsequent
low flip angle readout pulses (about 6$^{\circ}$)
\cite{Ardenkjaer-Larsen2003} with a repetition time $\tau$ between
120 s and 600 s. $^{13}$C steady state polarization $P_{\text{N}\infty}$
and the polarization time constant $T_{\text{pol}}$, describing
the progressive growth of the polarization, were derived by
fitting the build up curves to an expression that takes into
account the reduction of the $^{13}$C signal amplitude induced by
the readout pulses \cite{guilhem}. In the absence of MW
irradiation the same sequence was used to measure $^{13}$C
$T_1$ ($T_{\text{1C}}$) by following the build up of the $^{13}$C
NMR signal to the thermal equilibrium value after RF saturation.
Alternatively, $T_{\text{1C}}$ was derived from the decay of the
steady state polarization to thermal equilibrium after switching
off the MW, again measured by using a low flip angle (about
6$^{\circ}$) sequence (Fig. \ref{fgr:buildup}). The $^{13}$C NMR
signal decay was fit to the following expression
\begin{equation}\label{eq1}
 M(t)=M_{\infty} \exp\left[ -\left( \frac{t}{T_{\text{1C}}}-\frac{t \log(\cos \alpha)}{\tau}\right) \right] + M_{0},
\end{equation}
with $M_{\infty}$ the steady state $^{13}$C magnetization under MW
irradiation, $\alpha$ the flip angle, $\tau$ the repetition time
(300 s - 800 s) and $M_{0}$ the $^{13}$C thermal equilibrium
magnetization. The logarithmic term in Eq. \ref{eq1} takes into
account the artificial reduction of the NMR signal induced by the
readout pulses.
\begin{figure}[bottom]
\centering
  \includegraphics[height=6.8cm]{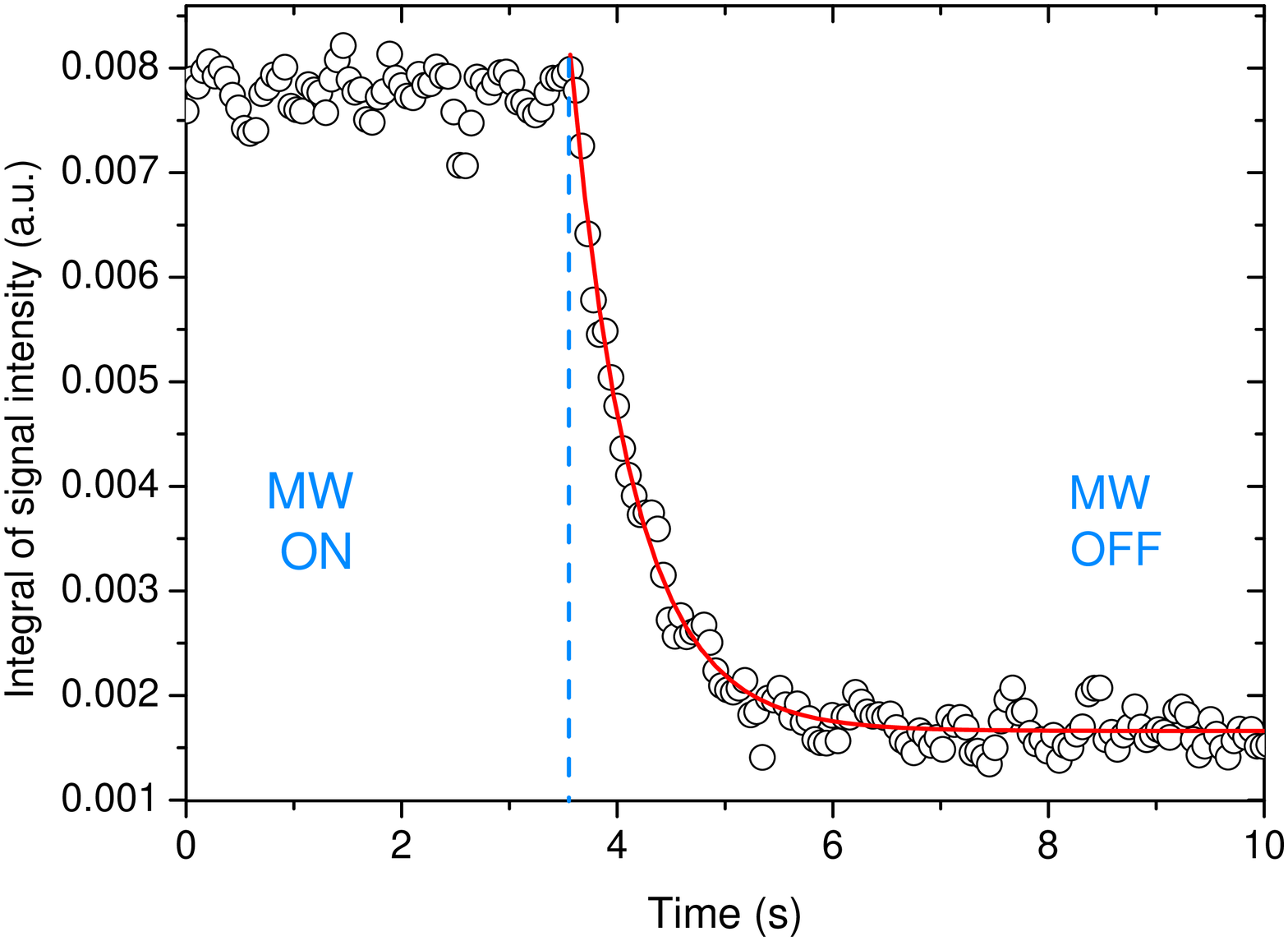}
  \caption{Integral of the imaginary part of the $^{13}$C signal as a function of time in PA15 at 2 K by means of a low flip angle acquisition scheme with $\alpha= 3^{\circ}$ and $\tau=30$ ms. In this experiment the MW were switched off 3.5 s after the sequence start. Data have been corrected by the artificial decay induced by the application of the readout pulses. The points collected after switching off MW could be fit to a simple exponential decay $y(t)=A \exp(-t/T_{\text{1e}})$ (red curve).}
  \label{fgr:t1edecay}
\end{figure}

The electron spin-lattice relaxation time $T_{\text{1e}}$ was derived
indirectly by observing the effect of the time evolution of
electron spin polarization on the NMR paramagnetic shift, and
hence on the NMR signal, after the MW were turned off. In
particular, after RF saturation the sample was polarized under MW
irradiation for about 10-15 minutes. This time is enough for the
electrons to reach steady state saturation and, additionally, to
increase the $^{13}$C signal-to-noise ratio significantly without
having to wait the long time required to reach $P_{\text{N}\infty}$. Subsequently, a low flip angle
acquisition sequence was started, with 3$^{\circ}$ flip angles and
with time delay between consecutive FID acquisitions between 15 ms
and 100 ms. Around 3-9 s after the beginning of the sequence, MW
were switched off and the $^{13}$C NMR signal was followed for few seconds.
During this time window, since the nuclear relaxation is orders of magnitude slower than the electronic one, the paramagnetic shift of the $^{13}$C
NMR line $\Delta \omega_\text{0}$ is found to vary proportionally to $P_{\text{e}}
(t)\propto \exp(-t/T_{\text{1e}})$ and to finally reach a plateau. The variation of  $\Delta \omega_\text{0}$
correspondingly implies a modification of the shape of the NMR
signal. J\'{o}hannesson et al. \cite{Johannesson2009} have
described a detailed procedure which allows to analyze the NMR
signal shape and to quantify the $^{13}$C NMR line shift. However
as long as the precise determination of the line shift is not
concerned, more workable approaches can be adopted to estimate $T_{\text{1e}}$.
In this work $T_{\text{1e}}$ was extracted by fitting the decay of the
integral of the imaginary part of the $^{13}$C signal $I(t)$,
obtained after switching MW off, to a simple exponential decay $A
\exp(-t/T_{\text{1e}})$ (Fig. \ref{fgr:t1edecay}). Further details on the
procedure used to derive $T_{\text{1e}}$ are given in Appendix 6.2, where a complete description of the NMR shift time dependence in presence of both electron polarization dynamics and the nuclear polarization dynamics is also provided and discussed.

\section{Experimental Results}
\begin{figure}[h!]
\centering
  \includegraphics[height=6.6cm]{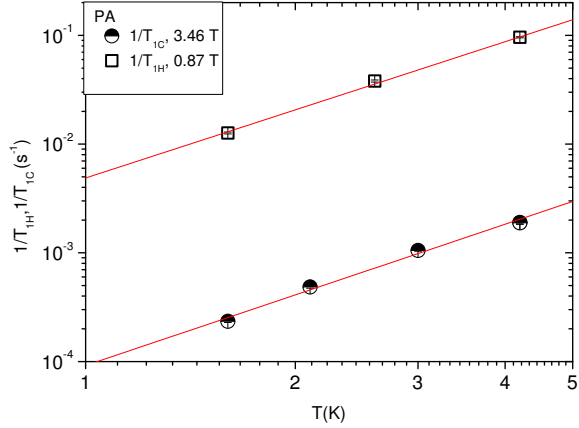}
  \caption{Log-log plot of $1/T_{\text{1H}}(T)$ (squares) and  $1/T_{\text{1C}}(T)$ (circles) in PA below 4.2 K. The red lines are fits to the power law $y(T)=a T^{b}$, yielding $a=9.19 \pm 1.1\cdot 10^{-5}$ and $b=2.16 \pm 0.11$ for $1/T_{\text{1C}}(T)$ and $a=4.88 \pm 0.44\cdot 10^{-3}$ and $b=2.08 \pm 0.07$ for $1/T_{\text{1H}}(T)$.}
  \label{fgr:t1n13c1h}
\end{figure}

\begin{figure}[h!]
\centering
  \includegraphics[height=7.2cm]{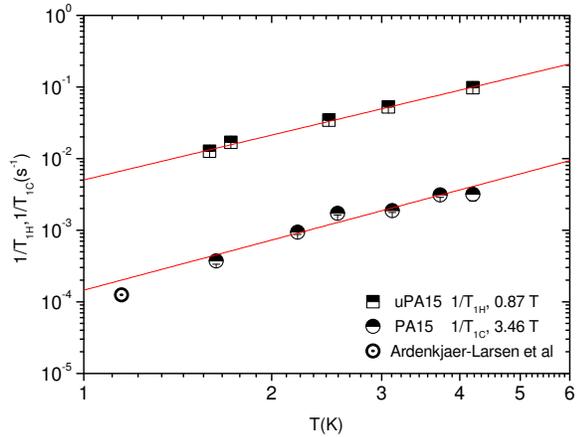}
  \caption{Log-log plot of $1/T_{\text{1H}}(T)$ (squares)  in uPA15 and of $1/T_{\text{1C}}(T)$ in PA15 (circles) below 4.2 K. The red lines are fits to the power law $y(T)=a T^{b}$, yielding $a=1.45 \pm 0.46\cdot 10^{-4}$ and $b=2.32 \pm 0.3$ for $1/T_{\text{1C}}(T)$ and $a=5.03 \pm 0.29\cdot 10^{-3}$ and $b=2.08 \pm 0.06$ for $1/T_{\text{1H}}(T)$.}
  \label{fgr:t1ncompariison}
\end{figure}

\begin{figure}[h!]
\centering
  \includegraphics[height=7cm]{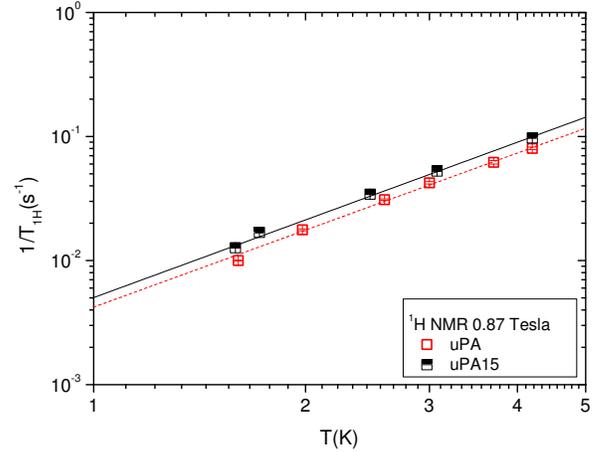}
  \caption{Log-log plot of $1/T_{\text{1H}}(T)$ measured in uPA15 (black squares) and in uPA (red squares) below 4.2 K. Both the black and the red lines are fits to the power law $y(T)=a T^{b}$. The black line is the same data fit of $1/T_{\text{1H}}(T)$ in PA15 reported in Fig. \ref{fgr:t1ncompariison}, while the red line has been obtained with the parameters $a=4.21 \pm 0.31\cdot 10^{-3}$ and $b=2.06 \pm 0.06$.}
  \label{fgr:t1n-1h}
\end{figure}

A different behaviour of $1/T_{\text{1H}}$ and $1/T_{\text{1C}}$ data was
observed by changing the cooling rate of PA and PA15 below 300 K.
The cooling rate dependence of $1/T_{\text{1C}}$ and the cooling
procedures are described in detail in Appendix 6.1. All the relaxation measurements presented in the following
were performed after flash freezing
the samples in liquid helium. The $T$ dependence of $1/T_{\text{1H}}$ and $1/T_{\text{1C}}$,
derived as explained in Section 2, are shown in Figs.
\ref{fgr:t1n13c1h}, \ref{fgr:t1ncompariison} and \ref{fgr:t1n-1h}.
The data in Fig. \ref{fgr:t1n13c1h} and \ref{fgr:t1ncompariison},
measured in PA, PA15 and uPA15 by keeping the same $\omega _{L}$ for the two nuclei,
evidence that both $1/T_{\text{1H}}(T)$ and
$1/T_{\text{1C}}(T)$ roughly follow a similar power law $\sim T^{2}$ (Table
\ref{tbl:tab1}). It is further remarked that in PA15 the
prolongation of the fit curve of $1/T_{\text{1C}}(T)$ down to 1.15
K (Fig. \ref{fgr:t1ncompariison}) closely approaches the value
reported for an analogous sample in Ref. \cite{ganiso}.

\begin{figure}[t]
\centering
  \includegraphics[height=6.6cm]{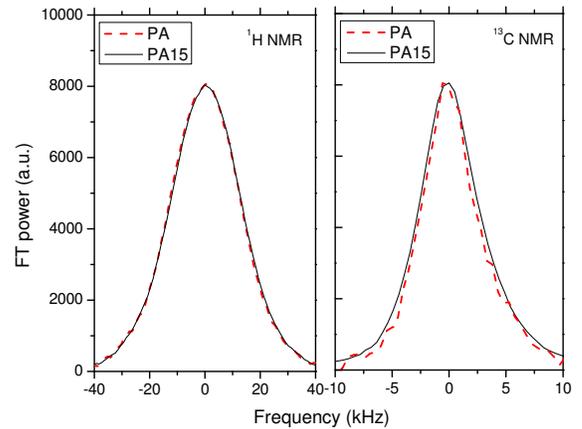}
  \caption{A comparison between the NMR lines of $^{1}$H in PA and PA15 (Left) and of $^{13}$C in PA and in PA15 (right) measured around 1.8 K. The frequencies in kHz are reported as shifts from the spectrometer frequency of 37.02 MHz. In order to compare the different NMR lines, all the plots have been centred around this same reference frequency.}
  \label{fgr:lines}
\end{figure}
\begin{figure}[h!]
\centering
  \includegraphics[height=6.6cm]{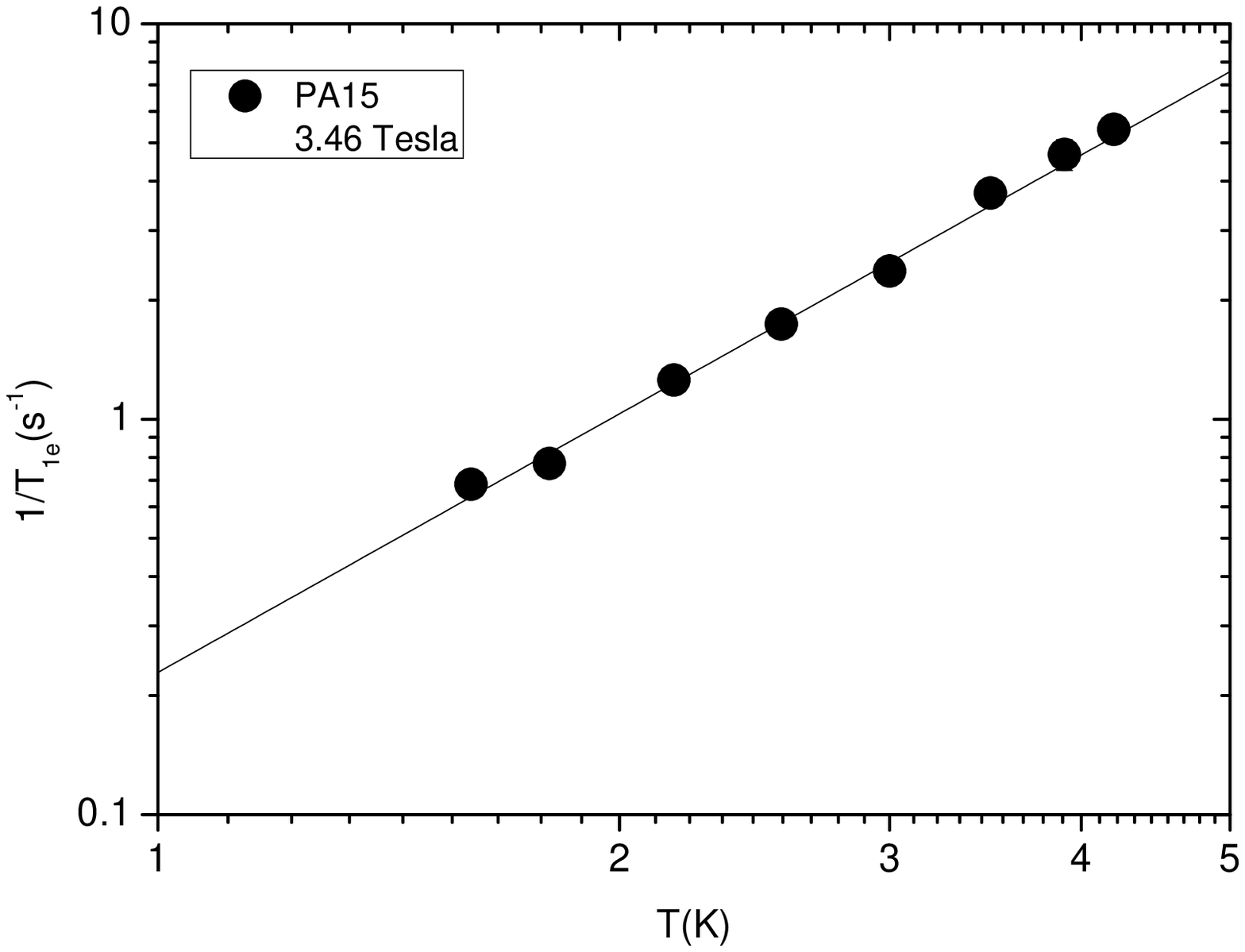}
  \caption{Log-log plot of $1/T_{\text{1e}}(T)$ in PA15 below 4.2 K. The black lines is a fit to the power law $y(T)=a T^{b}$, yielding $a=0.23 \pm 0.01\cdot 10^{-5}$ and $b=2.17 \pm 0.07$.}
  \label{fgr:t1et}
\end{figure}

In Fig. \ref{fgr:t1n-1h} the comparison between $1/T_{\text{1H}}(T)$
obtained in uPA15 and uPA is depicted. One can observe that also
in the radical free uPA sample $1/T_{\text{1H}}(T)$ follows a $\sim
T^{2}$ power law and, moreover, the comparison between the two
samples enlightens that the addition of 15 mM of OX063 radicals
yields only a minor enhancement of $1/T_{\text{1H}}(T)$ in the explored
$T$ range ($T_{\text{1H}}$(uPA15)/$T_{\text{1H}}$(uPA) $\simeq 1.2\div 1.3$). It is noticed that $1/T_{\text{1H}}(T)$ increases by a slightly lower amount of +15\% in the PA sample, in which also 1-$^{13}$C nuclei are present.

For the spin-spin relaxation times, we estimated
an almost $T$-independent $T_{\text{2C}}\approx 190$ $\mu$s and
$T_{\text{2H}}\approx 35$ $\mu$s in PA below 4.2 K. Also the width of the
NMR line was constant over the same $T$ range. The $^{1}$H lines of both PA and PA15 were fit to a Gaussian with a nearly equal Full Width at Half Maximum (FWHM) of $30.0 \pm 0.4$  kHz in PA and of $31.4  \pm  0.6$ kHz in PA15 between 1.6 K and 4.2 K (Fig. \ref{fgr:lines}). Differently,  $^{13}$C lines displayed a Voigtian lineshape with a FWHM of $5.9 \pm 0.1$ kHz in PA and of $6.1 \pm 0.1$ kHz in PA15 (Fig. \ref{fgr:lines}). The additional 200 Hz broadening in PA15 could be due to the coupling with electrons, however it is also of the order of the possible broadening due to the field inhomogeneity. 

Remarkably, also the electron spin-lattice relaxation rate $1/T_{\text{1e}}(T)$
measured after a flash freezing procedure (Fig. \ref{fgr:t1et})
could be fit to a $\sim T^{2}$ power law (Table \ref{tbl:tab1}). It can be noticed that
$T_{\text{1e}}$ increases progressively upon cooling until it reaches 1.5
s around 1.6 K, a value close to the one reported in the
literature at $T=1.2$ K \cite{ganiso}.
\begin{figure}[h!]
\centering
  \includegraphics[height=9cm]{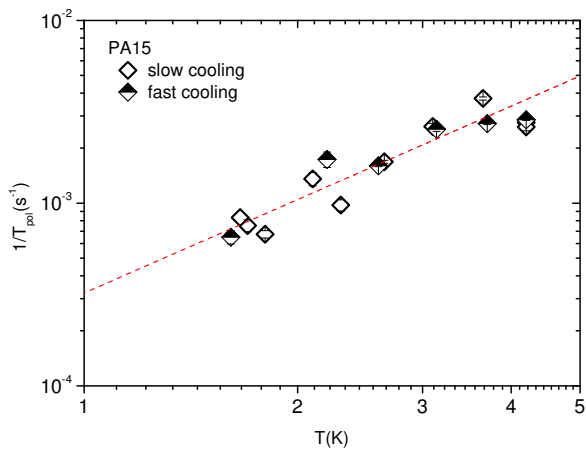}
  \caption{Log-log plot of $1/T_{\text{pol}}(T)$ in PA15 below 4.2 K measured after a slow cooling (white diamonds) and a fast cooling (black and white diamonds) procedure. The dashed line is a fit according to the power law $y(T)=a T^{b}$, yielding $a=3.02 \pm 0.27 \times 10^{-4}$ and $b=1.70\pm 0.18$}
  \label{fgr:Tpol}
\end{figure}

Now the $T$ dependence of the two characteristic DNP parameters
$T_{\text{pol}}$ and  $P_{\text{N}\infty}$ for the PA15 sample is
presented. As shown in Fig. \ref{fgr:Tpol}, $1/T_{\text{pol}}\sim
1/T_{\text{1C}}\approx 3\div 4 \cdot 10 ^{-3}$ s$^{-1}$ around 4.2 K and reduces significantly
on cooling. Correspondingly $T_{\text{pol}}$ reaches values around 1500 s for
$T \simeq 1.6$ K, much shorter than $T_{\text{1C}}\simeq
3000$ s at the same $T$. Moreover, $T_{\text{pol}}$ values at the
lowest $T$ of 1.6 K are close to the ones reported in the
literature at T$\simeq 1.2$ K \cite{macholl2010}. Also
$1/T_{\text{pol}}$ follows a power law $a T^{b}$ with $b\simeq
1.7$ (Table \ref{tbl:tab1}) in substantial agreement with literature papers
suggesting a proportionality between $T_{\text{pol}}(T)$ and $T_{\text{1e}}(T)$ \cite{walker,
shimon1} and reporting a divergence of $T_{\text{1C}}$ and
$T_{\text{pol}}$ (Fig.\ref{fgr:T1c-Tpol}) at very low $T$
\cite{walker, shimon1, shimon2, Ardenkjaer-Larsen2003, Johannesson2009, ganiso, jannin1, Lumata2011, Kurdzesau2008, plucktun}.
Nevertheless, to our knowledge the mechanism responsible for this phenomenon has not
been specifically addressed to date.

\begin{figure}[h!]
\centering
  \includegraphics[height=7.2cm]{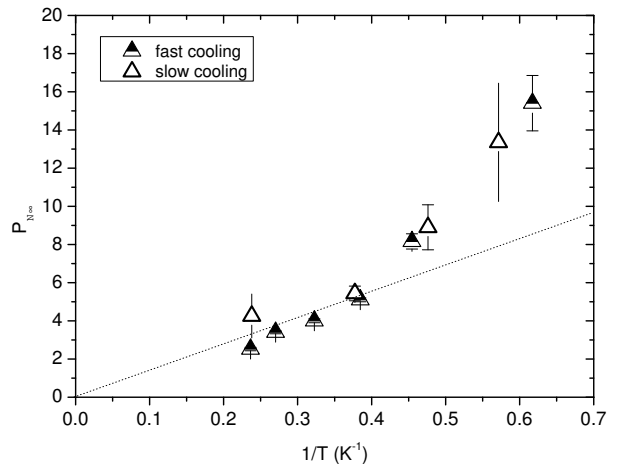}
  \caption{$P_{\text{N}\infty}$ as a function of $T$ in PA15 measured after a slow cooling (white triangles) and a fast cooling (black and white triangles) procedure. The error bars with caps have been estimated by the best fit of polarization builup curves, while the error bars without caps indicate the standard deviation for a series of repeated measurements.}
  \label{fgr:pol}
\end{figure}

The values of the steady state polarization $P_{\text{N}\infty}$ for the
PA15 sample, derived from the build up curves between 1.6 K and 4.2
K are reported in Fig. \ref{fgr:pol} as a function of $1/T$. $P_{\text{N}\infty}$ reaches already a sizeable value, around 3-4 \%,
at 4.2 K which raises up to 15.5 \% at 1.6 K, with a
linear trend at high $T$ that turns into a non linear bend at lower $T$ (for $1/T>0.4$
$K^{-1}$, i.e. $T<2.5$ $K$). These
values of $P_{\text{N}\infty}$, as well as the presence of the bending,
cannot be explained in the framework of the traditional Borghini model
\cite{borghini,abragorder} which predicts a polarization of $\sim
80\%$ at low $T$ and an opposite curvature for the bending.
Finally it is noted that, unlike nuclear spin-lattice relaxation data, both
$T_{\text{pol}}$ and  $P_{\text{N}\infty}$ do not depend on the cooling
rate.

\begin{table}[h]
\caption{\label{tbl:tab1}Fit results of the NMR and DNP measurements according to the law $y(T)=a T^{b}$ in PA samples at 3.46 Tesla}

\begin{tabular}{@{}llll}
\hline
Sample & Measurement & a (s$^{-1}\cdot K^{-b}$) & b\\
\hline
    PA   & $1/T_{\text{1C}}(T)$  & $9.19 \pm 1.11\times 10^{-5}$ &  $2.16\pm 0.11$\\
    PA   & $1/T_{\text{1H}}(T)$  & $4.88\pm 0.44\times 10^{-3}$ &  $2.08\pm 0.07$\\
    uPA  & $1/T_{\text{1H}}(T)$  & $4.21\pm 0.31\times 10^{-3}$ &  $2.06\pm 0.06$\\
    PA15 & $1/T_{\text{1C}}(T)$  & $1.45\pm 0.46\times 10^{-4}$ & $2.32\pm 0.3$  \\
    uPA15 & $1/T_{\text{1H}}(T)$ & $5.03\pm 0.29\times 10^{-3}$ & $2.08\pm 0.06$ \\
    PA15 & $1/T_{\text{1e}}(T)$  & $0.23 \pm 0.01$              & $2.17\pm 0.07$\\
    PA15 & $1/T_{\text{pol}}(T$) & $3.02 \pm 0.27 \times 10^{-4}$ & $1.70\pm 0.18$\\
\hline
\end{tabular}

\end{table}

\section{Discussion}

\subsection{Nuclear spin-lattice relaxation in pyruvic acid}

First we shall start considering the different contributions to
nuclear spin-lattice relaxation rate in PA. The main contribution
to $1/T_{\text{1H}}$ arises from the dipolar  $^1$H-$^1$H coupling while
only a minor contribution ($< 15 \%$) due to $^1$H-$^{13}$C
interaction is present, as it is evidenced by comparing $1/T_{\text{1H}}$
in PA and uPA samples. Thus in PA $1/T_{\text{1H}}$ can be expressed
as the sum of independent terms
\begin{equation}\label{eq2}
\left( \frac{1}{T_{\text{1H}}}\right)_{PA}=\left( \frac{1}{T_{\text{1}}}\right) _{^1H-^{1}H}+\left( \frac{1}{T_{\text{1}}}\right) _{^{13}C-^{1}H},
\end{equation}
where $(1/T_{\text{1}})_{^1H-^{1}H}$ sums up the contributions from intra
and intermolecular proton-proton dipolar interactions and
$(1/T_{\text{1}})_{^{13}C-^{1}H}$  originates from carbon-proton
intramolecular interactions. Accordingly, in uPA the second term
must be omitted. Similarly, in PA the $^1$H line broadening can be
ascribed mostly to $^1$H-$^1$H interactions. In fact, it is nearly
30 kHz in both PA and uPA, which demonstrates that the line
broadening due to the $^1$H-$^{13}$C interaction in the COOH group
is much smaller than the $^1$H-$^1$H coupling. Upon neglecting the
$^1$H-$^{13}$C interaction, from the second moment of the proton
line one can definitely estimate a mean square amplitude of the
dipolar field probed by protons $\sqrt{\langle\Delta
h^2\rangle}\simeq  7.6 \cdot 10 ^{-4}$ Tesla.

For $^{13}$C the relevant intra and intermolecular heteronuclear
dipolar interactions take place between the carboxyl $^{13}$C and
the methyl and hydroxyl protons. This coupling significantly
overcomes the homonuclear $^{13}$C-$^{13}$C one. In fact, as it is
described in detail in Appendix 6.3, according to the $1/T_{\text{1}}$
expression for homonuclear (Eq. \ref{eq16}) and heteronuclear (Eq.
\ref{eq17}) interactions, the ratio
\begin{equation}\label{eq3}
\frac{(T_{\text{1}})_{^1H-^{13}C}}{(T_{\text{1}})_{^{13}C-^{13}C}}\simeq\ 3
\left( \frac{\gamma_C}{\gamma_H}\right) ^2
 \frac{\left\langle\frac{1}{r^{6}_{CC}}\right\rangle}{\left\langle\frac{1}{r^{6}_{HC}}\right\rangle} \sim
10^{-4},
\end{equation}
where $\gamma_C$ is the $^{13}$C gyromagnetic ratio and $\gamma_H$ the $^{1}$H
gyromagnetic ratio, $\left\langle 1/r^{6}_{HC}\right\rangle$ indicates the average of the inverse sixth power of $^{1}$H-$^{13}$C distances $r_{HC}$ and $\left\langle 1/r^{6}_{CC}\right\rangle$ is the same quantity referred to $^{13}$C-$^{13}$C distances $r_{CC}$.
This difference is simply due to the fact that the average
intermolecular $r_{CC}$ is significantly larger
than the intramolecular $r_{HC}$ \footnote[1]{The
shortest $r_{HC}$ found in the COOH group is about
1.84 $\AA{}$ in the most abundant pyruvic acid conformer
\cite{Reva2001} while the intermolecular $r_{CC}$ should
rather be closer to 5.5 $\AA{}$, equal to twice the Van Der Waals
radius of the PA molecule.}. Additionally, in the case of $1/T_{\text{1C}}$ a further mechanism, involving the fluctuations of the
chemical shift tensor (CSA), should be considered and thus,
neglecting the weak homonuclear interactions, for PA one can write
\begin{equation}\label{eq4}
\left( \frac{1}{T_{\text{1C}}}\right)_{PA}=
(\frac{1}{T_{\text{1}}})_{^1H-^{13}C}+(\frac{1}{T_{\text{1}}})_{\text{CSA}},
\end{equation}
where $(1/T_{\text{1}})_{\text{CSA}}$ refers to the fluctuations of the
CS tensor. Accordingly, the $^{13}$C linewidth (5.9 kHz) cannot be
explained by considering  the $^1$H-$^{13}$C dipolar coupling
only. In this respect Macholl et al.\cite{macholl2010} report a
calculation of the 1-$^{13}$C CS tensor parameters in PA, that
retrieved the shielding anisotropy $\Delta \sigma \simeq 130$ ppm
and is thus responsible of a broadening at half height of the
1-$^{13}$C line of $\Delta \sigma \omega_{C}=4.8$ kHz at $H$=3.35 T and the Larmor frequency $\omega_C=\gamma_C H$. Only
the remaining broadening, at most 3.4 kHz, should thus be ascribed
to $^1$H-$^{13}$C couplings. This explains also why $^1$H-$^{13}$C
yields a negligible contribution to the $^1$H linewidth (less than 10
\% of the total width).

Now it is interesting to compare the ratio
$T_{\text{1C}}(T)/T_{\text{1H}}(T)\simeq 53$  (Fig.
\ref{fgr:t1n13c1h}) derived experimentally for the Larmor frequency
$\omega_H/2\pi=(\gamma_{H}/2\pi) H_{1}$ equal to $\omega_C/2\pi=(\gamma_C/2\pi) H_{2}=37.02$ MHz ($H_{1}=0.87$ Tesla and $H_{2}=3.46$ Tesla), with the one that can be estimated theoretically by considering the different contributions to the
nuclear spin-lattice relaxation.  One can start calculating that ratio under the assumption
that the spectral density $J(\omega)$ is the same for all the
relaxation contributions and specializing their expression taking
into account both the dipolar and CSA relaxation mechanisms (see
Eq.\ref{eq16} and \ref{eq17} in Appendix 6.3). Under that
assumption one finds an extremely good agreement between the
calculated $T_{\text{1C}}(T)/T_{\text{1H}}(T) \simeq 54$ (see
Appendix 6.3) and the experimental value, which indicates that
the $1/T_{\text{1}}$ models in Eq. \ref{eq2} for $^1$H and in Eq. \ref{eq4} for
$^{13}$C are likely correct. The fact that the same $J(\omega)$ describes
the relaxation mechanisms for the two nuclear species in PA will
be further discussed in the following Subsection.


\subsection{The role of the glassy matrix in the relaxation rates}

The nature of the excitations leading to the spin-lattice
relaxation is now analyzed. The common $T^2$ dependence of
$1/T_{\text{1H}}(T)$, $1/T_{\text{1C}}(T)$ and
$1/T_{\text{1e}}(T)$ strikingly points to the
presence of a common source of relaxation. In particular, while the
nuclear spin-lattice relaxation is dominated by the fluctuations of the dipolar interactions
with the other nuclei and with the electrons in PA15 (discussed in Subsection 4.3), the electron
spin-lattice relaxation of the diluted radicals is rather
induced by scattering with the vibrational modes. Therefore, the
lattice vibrations seem to be responsible both for spatial
modulation of dipolar couplings at the nuclear Larmor frequency
and for the excitation of electron spin transitions at the
electron Larmor frequency. The existence of a such a broad
spectral density of lattice excitations, even at liquid helium T,
should not surprise, since solid PA is an organic glass
\cite{Bohmer2001, Wiedersich}. Several physical properties of
glasses can be described by assuming a local lattice dynamics,
namely molecules or atoms can fluctuate among different
configurations having very similar energy minima, separated by a
barrier $\Delta E$. Upon increasing $T$ the correlation time of
these fluctuations can be described by an activated law
$\tau_{c}(T)=\tau_{0}\exp(\Delta E/T)$, with $\tau_0$ the
correlation time in the infinite $T$ limit.

For each activation barrier, $1/T_{\text{1}}$ can be simply described resorting to a
spectral density of the form
\begin{equation}\label{eq5}
\frac{1}{T_\text{1}}=\frac{\gamma^2\left\langle \Delta h^2_\perp\right\rangle}{2} J(\omega
_{L})=\frac{\gamma^2\left\langle \Delta h^2_\perp\right\rangle}{2}
\frac{2\tau_c}{1+\omega_L ^2 \tau_c^2} \,\,\, ,
\end{equation}
where $\left\langle \Delta h^2_\perp\right\rangle$ is the mean square amplitude of the
random fluctuating fields probed by the nuclei in the plane
perpendicular to the magnetic field. By considering different types of distribution functions $p(\Delta
E)$ for the energy barriers one typically finds a low-T power-law
behaviour with $1/T_1\sim T^{1+\alpha}$ ($0\leq \alpha\leq 1$). A
quadratic trend, as the one experimentally observed here, is
obtained for $p(\Delta E)\propto \Delta E$. Notably, the same
result was derived also  taking into account the thermally
activated dynamics in asymmetric double wells characterizing the
glasses \cite{Estalji,misra} and a recent implementation of the same
approach could explain also the quadratic $T$ dependence of
$1/T_{\text{1e}}$ observed at low $T$ in various amorphous
materials, including organic glasses \cite{Merunka2011}.

It is remarked that the magnetic field dependence of
$1/T_{\text{1C}}$ at 4.2 K shows that $1/T_{\text{1C}}\propto
1/\omega_L^2$, suggesting (see Eq.\ref{eq5}) that basically all
the lattice modes are characterized by low-frequency fluctuations
such that $\omega_L\tau_c\gg 1$. This observation is also
corroborated by the observation that $^{13}$C NMR linewidth is
T-independent in the explored T-range, indicating that
$2\pi\Delta\nu \tau_c\gg 1$. In the presence of such slow dynamics
one can consider the slow motions limit of Eq.\ref{eq5}, yielding
$1/T_{\text{1}}(T)\propto \left\langle 1/\tau_c (T)\right\rangle $, where $\left\langle 1/\tau_c \right\rangle$
represents an average correlation frequency of the fluctuations over
the distribution $p(\Delta E)$. $\left\langle 1/\tau_c (T)\right\rangle$ in uPA con be estimated by specializing $\left\langle \Delta h_\perp\right\rangle ^2$ in Eq. \ref{eq8} to
the case of $1/T_{\text{1}}$ driven by the dipolar interaction with like
spins (Eq \ref{eq16}), obtaining \cite{abragam}
 \begin{equation}\label{eq6}
\left( \frac{1}{T_{\text{1H}}}\right)_{uPA}=\frac{2}{5} (\frac{\mu_0}{4\pi})^2\frac{\gamma_H^4 \hbar^2 I(I+1)}{\omega_H ^2}\left\langle\frac{1}{r^{6}_{HH}}\right\rangle
 \left\langle\frac{2}{\tau_c}\right\rangle,
 \end{equation}
where $I=1/2$ is the proton spin,  $r_{HH}$ the inter-proton distance and
$\omega_H/2\pi= 37.02$ MHz. Since the
temperature dependence of $1/T_{\text{1}}$ is entirely contained in
$\left\langle 1/\tau_c (T)\right\rangle$, then $\left\langle 1/\tau_c (T)\right\rangle= 1/T_{\text{1}}(T) 1/C=
A_{1}*T^{B_{1}} *1/C =A T^B$, where C  is a factor including the
temperature independent parameters  in Eq .\ref{eq6} and $A_{1}$
and $B_{1}$ are the fit parameters of $1/T_{\text{1}}$ in uPA reported in
Table 1. Then $A=A_{1}/C$ and  $B=B_{1}$ and considering a mean
dipolar field of $7.6 \cdot 10 ^{-4}$ Tesla, as estimated from $^1$H NMR linewidth,
one finds $A\simeq 6.7\times 10^{3}$ s$^{-1} \cdot$K$^{-B}$ and
$B\simeq 2.06$. The same procedure can be applied also to evaluate
$\left\langle 1/\tau_c (T)\right\rangle$ from $^{13}$C data in PA, by considering Eq.
\ref{eq4}, and Eq. \ref{eq17}, Eq. \ref{eq18} in the slow motion
regime (see also Appendix 6.3 and \cite{abragam,Kowalewski})
 \begin{equation}
 \begin{split}\label{eq7}
 \left( \frac{1}{T_{\text{1C}}}\right) _{PA}= & \frac{2}{15} \left( \frac{\mu_0}{4\pi}\right) ^2 \frac{754}{225} \gamma_C^2 \gamma_H^2  \hbar^2 I(I+1)\left\langle\frac{1}{r^{6}_{HC}}\right\rangle \frac{1}{\omega_C ^2}\left\langle\frac{1}{\tau_c}\right\rangle + \\ & + \frac{2}{15}\Delta \sigma ^{2}\left\langle\frac{1}{\tau_c}\right\rangle,
 \end{split}
 \end{equation}
where $\omega_C/2\pi= 37.02$ MHz. Using the
value of $\left\langle \Delta \omega_{\perp \text{CH}} ^2\right\rangle  =
2/15(\mu_0/4\pi)^2 \gamma_C^2 \gamma_H^2
\hbar^2 I(I+1)\left\langle1/r^{6}_{HC}\right\rangle = 232.5$ $(\text{krad/s})^2$ obtained from the $^{13}$C linewidth analysis reported in literature, one finds $A\simeq 5.5\times 10^{3}$ s$^{-1}
\cdot$K$^{-B}$ and $B\simeq 2.16$, a value very close to the one
calculated for protons. Again, this results do confirm that the
leading modulation source for all the interactions probed by
nuclei, both dipolar and due to CSA, is the glassy dynamics of the
PA matrix.

\subsection{Nuclear spin-lattice relaxation in the presence of radicals}

Upon doping PA with trityl radicals, the $1/T_{\text{1}}$ analysis has to be
modified in order to include also a relaxation term due to the
coupling with the radical electron spins. Hence in PA15 Eqs.
\ref{eq2} and \ref{eq4} are modified as
\begin{equation}\label{eq8}
\left( \frac{1}{T_{\text{1H}}}\right)_{PA15}=(\frac{1}{T_{\text{1}}})_{^1H-^{1}H}+(\frac{1}{T_{\text{1}}})_{^{13}C-^{1}H}+(\frac{1}{T_{\text{1H}}})_{el}
\end{equation}
and
\begin{equation}\label{eq9}
\left( \frac{1}{T_{\text{1C}}}\right)_{PA15} =
(\frac{1}{T_{\text{1}}})_{^1H-^{13}C}+(\frac{1}{T_{\text{1}}})_{\text{CSA}}+(\frac{1}{T_{\text{1C}}})_{el},
\end{equation}
where $(1/T_{\text{1}})_{el}$ is the contribution due to the hyperfine
coupling between the nuclei and the radical electron spins. As it
is evident from Fig. \ref{fgr:t1n-1h}, the most relevant
contributions to $1/T_{\text{1H}}$ in PA15 is still due to the
nucleus-nucleus dipolar interaction. Doping yields only a slight
increment of $1/T_{\text{1H}}$ in PA15 with respect to PA, of the
order of +$20\div 30$\%. On the other hand, according to Table
\ref{tbl:tab1}, one observes that
$T_{\text{1C}}(PA)/T_{\text{1C}}(PA15)\simeq 1.6$, meaning that
the electron contribution to $T_{\text{1C}}$ is about $2\div 3$ times the
one to $T_{\text{1H}}$.

Within a simplified model one could start analyzing these results
by simply considering an electron contribution to the longitudinal relaxation of the $^{13}C$ nuclei driven by the lattice dynamics only.
As illustrated in Appendix 6.3, the average ratio
$T_{\text{1C}}(T)/T_{\text{1H}}(T)\simeq 28 \pm 3$ in PA15 (Fig.
\ref{fgr:t1ncompariison}) can be reasonably explained within this framework. 
However, this simplified model suffers from several limitations being based on certain approximations for the magnitude of the fluctuating fields and of $\tau_c$. For example, it does not take into account the effect of spin diffusion. On the other hand, as it will be shown in Sect. 4.4, $(1/T_{\text{1C}})_{el}$ can be quantitatively explained without adjustable parameters in the framework of the TM approach. 

Still, $(1/T_{\text{1H}})_{el}$ cannot be ascribed to TM, because 
$^1$H nuclei are characterized by $\omega_L$ larger than the electron spin resonance (ESR)
linewidth. Thus, in the case of $^1H$ nuclei, $(1/T_{\text{1H}})_{el}$ should be due exclusively to the
modulation of the electron dipolar field at the nucleus caused by the glassy dynamics. Accordingly one can estimate
$\left\langle 1/\tau_c (T)\right\rangle $ associated with $1/T_{\text{1H}}$ due to radicals, given
by $(1/T_{\text{1H}})_{el}=
(1/T_{\text{1H}})_{\text{uPA15}}-(1/T_{\text{1H}})_{\text{uPA}}$,
by applying the same approach adopted in the previous Subsection 4.2
and taking into account the appropriate magnitude
of the hyperfine coupling. In particular, in the slow motion regime one can
write \cite{abragam}
 \begin{equation}\label{eq10}
 \left( \frac{1}{T_{\text{1H}}}\right) _{el}=\frac{2}{5} \left( \frac{\mu_0}{4\pi}\right) ^2\frac{\gamma_H^2 \gamma_e^2 \hbar^2 S(S+1)}{\omega_H ^2}\left\langle\frac{1}{r^{6}_{eH}}\right\rangle
\left\langle \frac{1}{\tau_c}\right\rangle,
 \end{equation}
where S is the electron spin, $\gamma_e$ the electron gyromagnetic
ratio and $r_{eH}$ the electron-proton distance. By assuming an
average hyperfine field of $\sqrt{\left\langle \Delta h_{eH} ^2\right\rangle} =8.2\cdot 10 ^{-4}$
Tesla at the $^1$H site \footnote[2]{One can consider the
hyperfine interaction between the protons and the neighbouring
radical electron spins in a region comprised between an inner
sphere having the radius of the radical $R_{1}=5.8$ \AA, and an
outer sphere with radius $R_{2}=(3\cdot0.74/4\pi c)^{1/3}= 26.9$
\AA , corresponding to half of the average distance among the
radicals in PA15. In this case, on neglecting the effect of spin diffusion, $\left\langle  1/r^{6}_{eH}
\right\rangle =1/{(R_{1}R_{2})}^3$, yields an estimate of the average hyperfine field of
$\sqrt{\left\langle \Delta h_{eH} ^2\right\rangle}  =  (\mu_0/4\pi) \gamma_e
\hbar \left[S(S+1)\left\langle 1/r^{6}_{eH}\right\rangle\right]^{1/2} = 8.2\cdot 10^{-4}$ Tesla at
the $^1$H site.}, very close to the one produced by the other
nearby protons, one finds $\left\langle 1/\tau_c (T)\right\rangle \simeq A T^{B}$ with $A\simeq
2.3\times 10^{3}$ s$^{-1} \cdot$K$^{-B}$ and $B\simeq 2.16$.
Remarkably, the $\left\langle 1/\tau_c(T)\right\rangle $ describing the fluctuations
leading to the relaxation term $(1/T_{\text{1H}})_{el}$ is close
to the one describing the glassy dynamics probed by
$(1/T_{\text{1H}})_{PA}$. Then one can conclude that in presence
of radicals the $^1$H relaxation involves the modulation of the
field generated by the paramagnetic radicals, driven by the
lattice glassy dynamics. Relaxation processes for
$(1/T_{\text{1H}})_{el}$  driven by electron spin flips can be
disregarded since they should be characterized by a fluctuation
frequency $1/T_{\text{1e}}$ much smaller than the one of the
glassy dynamics.
\begin{figure}[t]
\centering
  \includegraphics[height=7.2cm]{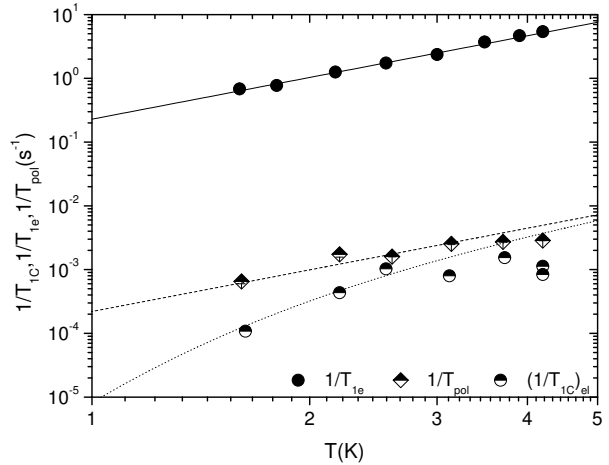}
  \caption{Comparison among the $T$ dependence of $1/T_{\text{1e}}$ (black dots), $1/T_{\text{pol}}$ (black and white diamonds) and $(1/T_{\text{1C}})_{el}$ (black and white circles) in PA15 below 4.2 K. The solid line is the fit of $1/T_{\text{1e}}$ according to the power law $y(T)=a T^{b}$, yielding the parameters reported in Tab. \ref{tbl:tab1}. The dashed line shows the function $(N_{e}/N_{N})1/T_{\text{1e}}$, while the dotted line gives  $(N_{e}/N_{N})1/T_{\text{1e}}[1-P_{0}(T)^2]$.}
  \label{fgr:T1c-Tpol}
\end{figure}

\subsection{The effect of Thermal Mixing in $^{13}$C spin-lattice relaxation and the electron spin-lattice relaxation}

As it was pointed out in the previous paragraph for $^{13}$C
nuclei a different scenario must be considered. Since $\omega_C$ is smaller than the ESR
linewidth, the $^{13}$C and the electron dipolar reservoirs are in
TM.  Within the TM process which governs $^{13}$C electron-nucleus
relaxation one has $(1/T_{\text{1C}}(T))_{el}= 1/T_{\text{1e}}(T)
(N_{e}/N_{n}) [1-P_{0}(T)^{2}]$. The ratio $N_{e}/N_{n}$ between
the radical and $^{13}$C concentrations definitely sets the order
of magnitude of $(1/T_{\text{1C}}(T))_{el}$ with respect to
$1/T_{\text{1e}}(T)$ and encloses a precise physical meaning: the
three body mechanism originating TM, involving two electron spins
and one nuclear spin, can flip one of the $N_{n}$ nuclear spins,
as long as one of the $N_{e}$ electrons relaxes to thermal
equilibrium. The electron contribution to $^{13}$C spin-lattice relaxation
$(1/T_{\text{1C}}(T))_{el}=
(1/T_{\text{1C}}(T))_{PA15}-(1/T_{\text{1C}}(T))_{PA}$ derived
from the experimental data sets is shown in Fig.
\ref{fgr:T1c-Tpol} (black and white circles). It is remarkable to
notice that  below 4 K $(1/T_{\text{1C}}(T))_{el}$ data
quantitatively follow the trend of the dotted function
$1/T_{\text{1e}}(T) (N_{e}/N_{n}) [1-P_{0}(T)^{2}]$, derived from
the fit function of the experimental $1/T_{\text{1e}}(T)$ data
(Table \ref{tbl:tab1}), with no adjustable parameter. This is a
clear evidence that indeed the $^{13}$C spin ensemble and the
electron dipolar reservoirs are strongly coupled in the TM, at
least for $T< 4$ K. This important observation is further
supported by the $T$ dependence of $1/T_{\text{pol}}(T)$, as
discussed in the following subsection.

The functional $T$ dependence $1/T_{\text{1e}}\propto T^{2.2}$ has been ascribed to
scattering with the glassy modes, although a sizeable dependence
of the $1/T_{\text{1e}}$ magnitude on $c$ is also expected from
previous investigations \cite{Johannesson2009, macholl2010}. Even
if literature data on $1/T_{\text{1e}}$  measured in different
experimental setup and at different fields are partially contradictory
\cite{walker,lumataepr,macholl2010, ganiso, Johannesson2009,
hess}, both the measurements at 1.2 K shown in \cite{macholl2010}
and our measurements in the 1.6-4.2 K range at higher $c$
(unpublished data), consistently evidence a linear dependence
$1/T_{\text{1e}}\propto c$ for trityl radicals in PA.
Thus, electron spin-lattice relaxation in trityl doped PA can be written as
$1/T_{\text{1e}}(T)=(1/T_{\text{1e}})_{g}(T)+ \Omega c$, where
$(1/T_{\text{1e}})_{g}(T)$ is the term linked to the glassy
dynamics and $\Omega$ is a phenomenological weakly $T$-dependent
parameter. According to \cite{macholl2010}
$(1/T_{\text{1e}})_{(PA15)}/(1/T_{\text{1e}})_{g}\sim 2\div3$ for
$H=3.35$ Tesla, indicating that for $c=15$ mM the contribution
due to dipole-dipole interactions among radicals significantly
overcomes the one originated by the scattering with the glassy
modes and that $1/T_{\text{1e}}(T)$ should
weakly depend on the cooling rate, at variance with nuclear $1/T_{\text{1}}$.

Overall, from the above considerations a clear scenario emerges.
In PA15 the $1/T_{\text{1H}}$ processes show a $T$ dependence which is
uniquely determined by the properties of the glassy matrix. On the
other hand, the dominant relaxation mechanism for $^{13}$C rather
involves the coupling of the nuclei to the electron dipolar
reservoir through TM. Notably, due to the glassy dynamics which
characterizes PA, the magnitude of $T_{\text{1C}}$ and $T_{\text{1H}}$ and their
T dependence can possibly vary among samples containing the same
radicals admixed to different molecular substrates or among
samples prepared, treated and cooled with different methods, which
yield to a different glassy dynamics at low T. The contribution of
the glassy modes to the electron spin-lattice relaxation can also justify some
variability of $T_{\text{1e}}$ data measured by different groups in
different conditions \cite{walker,lumataepr,macholl2010, ganiso,
Johannesson2009, hess}, even if the dominant electron-electron
dipolar relaxation mechanism yields to a $1/T_{\text{1e}}(T)$ which
scarcely depends on the cooling rate.

\subsection{Dynamical nuclear polarization}

As shown in Fig. \ref{fgr:Tpol} a nearly quadratic $T$ dependence
is found for $^{13}$C $1/T_{\text{pol}}$, the DNP build up rate.
The most recent models describing DNP through TM \cite{colombo1,colombo2,colombo3}
have shown how the nuclear polarization under MW irradiation can be deeply influenced
by several parameters such as $T_{\text{1e}}$, $T_{\text{ISS}}$,
the contact time between the nuclear Zeeman reservoir and the
electron dipolar reservoir, as well as by the dissipative spin
diffusion among electrons and the degree of saturation by the MW.
The behaviour of $T_{\text{pol}}$ depends on the ratio
$T_{\text{ISS}}/T_{\text{1e}}$. In particular, in presence of
nuclear leakage, for $T_{\text{ISS}}/T_{\text{1e}}\sim 1$
(poor contact between electrons and nuclei), polarization
levels much lower and $T_{\text{pol}}$ values longer than those
derived here are expected. On the other hand, for
$T_{\text{ISS}}/T_{\text{1e}}\ll 1$, the polarization should
increase and $T_{\text{pol}}$ should shorten and depend on
$T_{\text{1e}}$. Indeed in PA15 $1/T_{\text{pol}}$ has the same
$T^{2}$ dependence of $1/T_{\text{1e}}$ and its order of magnitude
matches quantitatively the functional dependence of
$(N_{e}/N_{N}) 1/T_{\text{1e}}$ below 4 K (dashed curve in Fig.
\ref{fgr:T1c-Tpol}). Thus, it is tempting to state that
a very efficient contact is actually attained. Remarkably, below 4
K the ratio $T_{\text{pol}}(T)/T_{\text{1C}}(T)_{el}$ behaves
as $\left[ 1-P_{0}(T)^{2}\right]$. This can happen only if both
polarization under MW irradiation and $T_{\text{1C}}$ relaxation
proceed through the same TM processes.

Both the polarization and the relaxation time of PA15 are
consistent with the TM regime, whereas  the experimental
$P_{\text{N}\infty}$ (Fig. \ref{fgr:pol}) is substantially smaller than the one
predicted by the traditional Borghini model and a mechanism of
DNP-dissipation should be identified. The dissipation inside the
nuclear reservoir via $^{13}$C spin-lattice relaxation is expected to be irrelevant because
the cooling procedure affects the value of $T_{\text{1C}}$ of PA,
but not $P_{\text{N}\infty}$ and only weakly $T_{\text{1e}}$, which
mainly depends on $c$. Thus the dissipation mechanism does likely affect directly
the electron reservoir and can be induced either by a limited microwave
power \cite{samisat, colombo2} or by the presence of dissipative
processes in the spectral diffusion as discussed in
\cite{colombo3}. At this stage we cannot exclude one of the two
mechanisms, or a combination of the two. Our simulations of the
rate equation model introduced in \cite{colombo1,colombo2,colombo3} show that  the bending behaviour
observed in Fig. \ref{fgr:pol} is consistent with both mechanisms.

Now, the quality of the electron-nucleus contact in PA seemingly
evolves on raising $T$ up to 4 K. In fact, in Fig. \ref{fgr:T1c-Tpol}
a deviation from the dotted and the dashed curves, tracing the
good contact trend, is noted around $T\sim 4$ K for $1/T_{\text{1C}}$ and
$1/T_{\text{pol}}$. Both time constants become longer than expected, which
likely indicates a substantial degradation of the electron-nucleus
contact. We possibly ascribe the worsening of the electron-nucleus
contact on raising $T$ to the shortening of $T_{\text{1e}}$. Infact, for
$T\approx 4$ K, $T_{\text{1e}}$ is around 200 ms, a value close to the
effective spin diffusion time \cite{blumberg} $(N_n/N_e) T_{\text{2C}}= 190$ ms, which in
turn determines the effective order of magnitude of $T_{\text{ISS}}$
between the electron dipolar reservoir and the whole nuclear spin
ensemble. Then, as explained before, for $T$ = 4 K the threshold of
the bad contact regime $T_{\text{1e}}\approx T_{\text{ISS}}$ is matched, the
polarization bottleneck becomes $T_{\text{ISS}}$ and $T_{\text{1C}}$ and
$T_{\text{pol}}$ become longer than expected in the good contact scenario. On
the other hand, it should be noted that in the explored $T$ range
any modulation of electron-nucleus coupling by the glassy dynamics
looks definitely ineffective. Infact, since in PA15 the modulation
of the electron-nucleus distances occurs over the frequency scales of
the glassy dynamics,  $10^{4}$ s$ < \langle 1/\tau_{c}\rangle
< 10^{5}$ s, and considering that the magnitude of the dipolar
coupling of the electron spins with the nearby nuclei $\Delta h$
is such that $\gamma_{^{13}C}\Delta h \geq \langle 1/\tau_{c}\rangle$, the TM mixing process is marginally affected by that dynamics, in agreement with the absence of any effect of the cooling history on
the DNP parameters.

Definitely one can conclude that in PA15 for $T < 4$ K TM occurs in a
good contact regime where $1/T_{\text{pol}}\propto 1/T_{\text{1e}}(T)$, while
for $T > 4$ K a bad contact regime is attained.
The $T$ dependence of $P_{\text{N}\infty}$ can be explained as well resorting to TM
models combined with dissipative mechanisms located in the electron spin system.

\section{Conclusions}

Through a series of nuclear spin-lattice relaxation measurements and DNP experiments, both in pure PA and
in radical doped PA, it was possible to evidence that several
microscopic parameters relevant for the understanding of the
dynamical nuclear polarization processes follow the same quadratic
$T$ dependence. This trend is interpreted in terms of the glassy dynamics which
characterize the PA at low $T$. Notably the $T$
dependence of the DNP build up time, of the electron contribution
to $1/T_{\text{1C}}$ and of the saturation polarization are found in
agreement with the TM regime with a very good thermal contact between
the nuclear and the electron non-Zeeman reservoirs between 1.6 K
and 4 K, where $1/T_{\text{pol}}\propto 1/T_{\text{1e}}(T)$. Above 4 K the TM
occurs through a less efficient contact, probably due to the
shortening of $T_{\text{1e}}$ which becomes of the order of $T_{\text{ISS}}$.
Definitively, this information gives an interesting feedback to the
latest theoretical developments, pointing out the relevance of the
electron spin relaxation processes, but more specifically claiming
a central role for the lattice excitations in determining the
ultimate DNP performances.

\section{Appendix}

\subsection{Dependence of the pyruvic acid dynamics on the cooling rate}

\begin{figure}[h!]
 \centering
   \includegraphics[height=6.8cm, bb= 200 0 380 380]{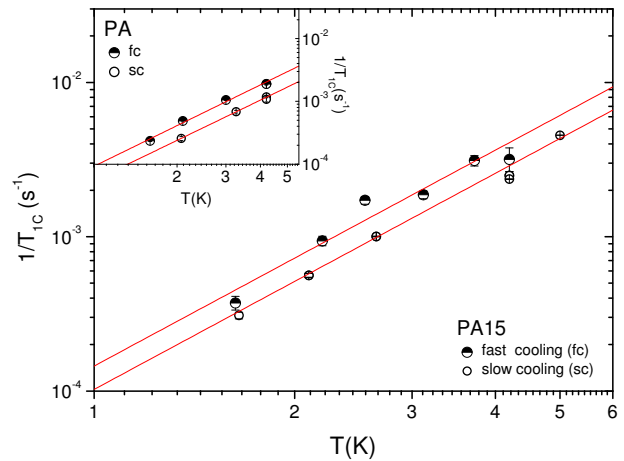}
   \caption{Log-log plot of $1/T_{\text{1C}}(T)$ measured in PA15 after a slow cooling (white circles) and a fast cooling procedure (black and white circles) below 4.2 K. Inset: Log-log plot of $1/T_{\text{1C}}(T)$ measured in PA after a slow cooling (white circles) and a fast cooling procedure (black and white circles) below 4.2 K. The red lines are fits to the power law $y(T)=a T^{b}$}.
   \label{fgr:cooling}
 \end{figure}

The dependence of the experimental results on the cooling method was verified by using two different procedures: {\em a})
a slow pre-cooling inside a bath cryostat from room $T$ to 150 K at -0.5 K/min, followed by a rapid cooling caused by the liquid helium fill; {\em b}) a flash freezing of the samples in liquid nitrogen, followed by immersion in liquid helium. Hereafter the first method will be indicated as slow cooling (sc), while the second as fast cooling (fc). Nuclear $1/T_{\text{1}}$ data showed a different behaviour with respect to the adopted cooling method in both PA and in PA15 (Fig. \ref{fgr:cooling}). This variation is likely due to a change in the matrix dynamics properties for different cooling procedures, which is typically observed in glasses. The comparison among $1/T_{\text{1C}}(T)$ in PA and PA15, points out that upon performing a fast cooling $1/T_{\text{1C}}(T)$ doubles in PA, while it only increases by a ratio of 1.5 in PA15. The reason of this difference is ascribed to the additional presence in PA15 of the relaxation term $(1/T_{\text{1C}})_{el}$ due to the thermal mixing with the electrons, which is rather unsensitive to the cooling rate, as $1/T_{\text{pol}}(T)$.

\subsection{Calculation of $T_{\text{1e}}$ from low flip angle acquisitions}

As explained, it is possible to quantify $T_{\text{1e}}$ by following the time variation of $\Delta \omega_\text{0}$. In PA15 the shift can be described by the sum $\Delta \omega_\text{0}=\Delta M_{\text{IS}}+\Delta M_{\text{II}}$, where $\Delta M_{\text{IS}}\propto P_{\text{e}}$ is generated by the hyperfine coupling between the nuclei and the electrons and $\Delta M_{\text{II}}\propto P_{\text{N}}$ by the dipolar nucleus-nucleus interactions \cite{abragorder}. Indeed, since for low $c$ $\Delta \omega_\text{0}$ is small (300 Hz at 1.2 K in PA15 \cite{ganiso,Johannesson2009}) its estimate from standard NMR line fits is critical. Conversely, it is rather advantageous to monitor it indirectly by analyzing the oscillations it induces in the NMR signal in the time domain.

When a line shift $\Delta \omega_\text{0}$ from the reference frequency of the NMR spectrometer $\omega_0$ is present, having moreover the NMR signal envelope $s(t)$ and  an arbitrary phase  $\phi$, in the domain of time (t) the imaginary component of the NMR signal $Im(t)$ has the form
\begin{equation}\label{eq11}
  Im(t)=x(t)\sin[(\Delta  \omega_0 t)+\phi].
\end{equation}

In particular, in the experiment performed to measure $T_{\text{1e}}$, also $\Delta \omega_\text{0}$ varies with time, but on the time scale of the whole NMR acquisition. The second time variable $t'$, triggered to the start of the experiment and with maximum value $N(\tau+ TD)$, where TD is the time domain of the single acquisition and $\tau$ the time delay between acquisitions, describes time evolution $\Delta\omega_{\text0} (t')$. Then Eq. \ref{eq11} more properly rules as:
\begin{equation}\label{eq12}
  Im(t,t')=x(t)\sin[(\Delta \omega_\text{0}(t')t)+\phi],
\end{equation}
and its integral as
\begin{equation}\label{eq13}
  I(t')=\int_{\tau1} ^{\tau2}Im(t,t')dt,
\end{equation}
in which the bounds $\tau1$ and $\tau2$ should be fixed in the interval in which $\arrowvert s(t)\arrowvert ^2 \neq 0$.
 \begin{figure}[h!]
 \centering
   \includegraphics[height=6.9cm]{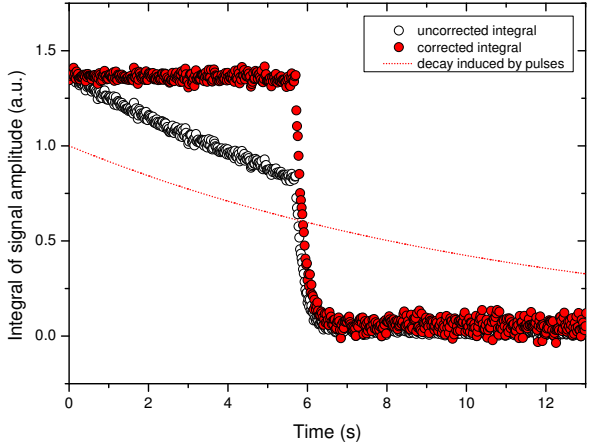}
   \caption{Simulation of the time dependence of the integral of the imaginary signal $I(t')$ in the sequence used for the $T_{\text{1e}}$ measurement at 4.2 K. The red dotted line shows the artificial decay induced by the read-pulses, the white circles show the uncorrected $I(t')$ as obtained by the simulation, the red circles correspond to $I(t')$ divided by the values of the red dotted line. The curves were simulated by setting $\Delta M_{\text{IS}}=1$ kHz and $\Delta M_{\text{IS}}/\Delta M_{\text{II}}=10$.}
   \label{fgr:simulation}
 \end{figure}

In this work the behaviour of $I(t')$ in Eq.\ref{eq12} was verified by means of a Python script on considering the complete shift dynamics $\Delta \omega_\text{0} (t')=\Delta M_{\text{II}}(t')+ \Delta M_{\text{IS}}(t')$. The NMR signal was modelled to a Gaussian decay, with $\sigma=105$ $\mu$s and an initial signal to noise ratio equal to 500. The simulation took into account also the reduction of the signal amplitude operated by the readout pulses (Fig.\ref{fgr:simulation}).  At time $ \overline{t'}$, when MW are switched off, for each $T$ $P_{\text{N}} (\overline{t'})/P_{\text{e}} (\overline{t'})$  was set to the maximum value $P_{\text{N}\infty}(T)/0.5$, calculated by taking into account that $P_{\text{N}} (T)(\overline{t'})<P_{\text{N}\infty}(T)$ and $P_{\text{e}} (\overline{t'})\geq 0.5$, the value of the residual electronic polarization $P^{res} _{\text{e}}$ expected for saturation at the frequency optimal for DNP \cite{Johannesson2009}.
For $t'>\overline{t'}$ considering both $T_{\text{1e}}$ and $T_\text{1N}$ spin-lattice relaxation processes and that $P_\text{N}$ is reduced on increasing $t'$ by the application of the read out pulses (see Eq. \ref{eq1}), the following laws were assumed to describe nuclear and electronic polarization
\begin{equation}\label{eq14}
  P_{\text{e}} (t')= (P^{res} _{\text{e}}-P_\text{E0})\exp(-t'/T_{\text{1e}})+P_\text{E0}
\end{equation}
\begin{equation}\label{eq15}
P_\text{N} (t')=(P_\text{N} (\overline{t'})-P_\text{N0})\exp\left[ -t'\left( \frac{1}{T_\text{1N}}-\frac{\log(\cos\alpha)}{\tau}\right) \right] +P_\text{N0},
\end{equation}
where $P_\text{N0}$ and $P_\text{E0}$ are the thermal equilibrium values for nuclear and electronic polarization respectively.
In Eq. \ref{eq15} $T_\text{1N}$ assumed the experimental values in Fig.\ref{fgr:t1ncompariison}, $\alpha=3 ^{\circ}$ and $\tau$ ranged from 15 ms at 4.2 K to 100 ms at 1.8 K. Remarkably, $P_\text{N} (t')$ has an effective relaxation rate which is driven mainly by the term $\exp(t'\log(\cos\alpha)/\tau))$, since $1/T_\text{1N}<<\log(\cos\alpha)/\tau$, and is increased sensibly by fast repetition (12 s for $\tau=19$ ms).
Eq. \ref{eq14} and \ref{eq15} were used to calculate $\Delta\omega_0(t')$ and then $I(t')$.
As expected, in spite of an increase of $P_\text{N} (\overline{t'})/P_{\text{e}} (\overline{t'})\simeq 0.26$ at 1.8 K, the simulation showed that imposing $T_{\text{1e}}\simeq 1$ s, $T_\text{1N}\simeq 1800$ s and $\tau=100$ ms, $I(t')$ is perfectly fit to a simple exponential decay until 60 s after $\overline{t'}$ with a decay constant of 1 s. At T=4.2 K for $T_{\text{1e}}\simeq 0.2$ s, $T_\text{1N}\simeq 400$ s, $\tau=19$ ms, until 12 s after $\overline{t'}$ the simple exponential fit of $I(t')$ led to a decay constant of 0.22 s, +11 \% with respect to the initial simulation parameter, only when increasing $P_\text{N} (\overline{t'})/P_{\text{e}} (\overline{t'})$ to 0.5, equal to 5 times the maximum reachable value at this T. Finally, the results didn't depend on the integration interval chosen to calculate $I(t')$.

$I(t')$ obtained by the experiment was divided by the expression in Eq \ref{eq1}, yielding a curve properly detrended by the artificial decay induced by pulses (Fig.\ref{fgr:t1edecay}). Eq. \ref{eq1} was considered valid also for $t'<\overline{t'}$ since for $\tau\ll T_{\text{pol}}$ the build up, occurring on times of the order of $T_{\text{pol}}\simeq 400$ s$\div 1200$ s is overwhelmed by the fast repetition of the read out pulses. Accordingly for all Ts and $t'>\overline{t'}$, experimental data of  $I(t')$ were suitably fit to a single exponential decay, after performing a smoothing procedure consisting in the unweighted averaging of 3 adjacent data points.

\subsection{Calculation of the $T_{\text{1C}}(T)/T_{\text{1H}}(T)$ ratio starting from model nuclear interactions}
In the following we are going to consider the contributions to spin-lattice relaxation of 1-$^{13}$C and $^{1}$H deriving from nuclear dipole-dipole interactions, CSA and hyperfine dipolar interactions.

The spin-lattice relaxation in case of homonuclear dipolar interactions between spins I will be described by \cite{abragam}
\begin{equation}\label{eq16}
(\frac{1}{T_{1}})_{I-I}=\frac{2}{5} \left( \frac{\mu_0}{4\pi}\right) ^2 \gamma_I ^4 \hbar^2 I(I+1)\left\langle \frac{1}{r^{6}_{II}}\right\rangle  [J(\omega_{I})+4J(2\omega_{I})],
\end{equation}
where $\gamma_I$ is gyromagnetic ratio of spins I, $\left\langle 1/r^{6}_{II}\right\rangle$ indicates the average of the inverse sixth power of I-I distances and $J(\omega _{I})$ the spectral density at the Larmor frequency $\omega _{I}=\gamma_I H$ of spins I.  The spin-lattice relaxation of spins I in case of interactions between spins I and S will be described by 
\begin{equation}
\begin{split}\label{eq17}
(\frac{1}{T_{1}})_{S-I} & =\frac{2}{15} \left( \frac{\mu_0}{4\pi}\right)^2\gamma_I ^2 \gamma_S ^2  \hbar^2 S(S+1) \left\langle\frac{1}{r^{6}_{SI}}\right\rangle \times \\
                        & \times [J(\omega_{I}-\omega_{S})+3J(\omega_{I})+6J(\omega_{I}+\omega_{S})],\\
\end{split}
\end{equation}
where $\gamma_S$ is gyromagnetic ratio of spins S, $\left\langle 1/r^{6}_{SI}\right\rangle$ is the average of the inverse sixth power of S-I distances and $\omega _{S}=\gamma_S H$ the Larmor frequency of spins S. Finally spin-lattice relaxation due to CSA at the 1-$^{13}$C site will be expressed as \cite{Kowalewski}
\begin{equation}\label{eq18}
(\frac{1}{T_{\text{1}}})_{CSA}=\frac{2}{15} \omega_{C}^{2} \Delta \sigma^{2} J(\omega_{C}),
\end{equation}
where $\Delta \sigma$ is the CSA tensor anisotropy.

In the following all the quantities referring to generic spins I ans S will be specialized to the case of protons (using subscript H), carbons (subscript C) and electrons (subscript e).
For homonuclear interactions among protons the mean square amplitude of the fluctuation frequencies of the dipolar field $\langle\Delta \omega_{\perp \text{HH}} ^2\rangle$ inducing the spin lattice relaxation is related to the powder line second moment  $\langle\Delta \omega_{\text{HH}} ^2\rangle$ to the relation
\begin{equation}\label{eq19}
\langle\Delta \omega_{\perp \text{HH}} ^2\rangle = \frac{2}{5}\left( \frac{\mu_0}{4\pi}\right)^2 \gamma_H ^4 \hbar^2 I(I+1) \left\langle\frac{1}{r^{6}_{HH}}\right\rangle = \frac{2}{3} \langle\Delta \omega_{\text{HH}} ^2\rangle,
\end{equation}
with spin I of protons equal to $1/2$, while for heteronuclear interactions the formula becomes
\begin{equation}\label{eq20}
\langle\Delta \omega_{\perp \text{HC}} ^2\rangle = \frac{2}{15}\left( \frac{\mu_0}{4\pi}\right)^2 \gamma_H ^2 \gamma_C ^2  \hbar^2 S(S+1) \left\langle\frac{1}{r^{6}_{HC}}\right\rangle = \frac{1}{2} \langle\Delta \omega_{\text{HC}} ^2\rangle,
\end{equation}
with spin S of carbons equal to $1/2$.

The important assumptions we make are:
\begin{list}{}{}
\item  1. $J(\omega)$ is the same for all the relaxation contributions and is the one related to the lattice dynamics.
\item  2. $J(\omega)$ is approximated to the slow motion regime form considering an average correlation frequency, yielding $J(\omega_{L})\simeq \frac{1}{\omega_{L}}^{2}\langle \frac{1}{\tau _{c}}\rangle$
\item 3. The amplitude of the local fluctuating fields probed by the nuclei is the one estimated from the linewidth analysis of  1-$^{13}$C and $^{1}$H.
\end{list}

From the literature data \cite{macholl2010}  it is possible to extract $\langle\Delta \omega_{\text{HC}} ^2\rangle= \langle\Delta \omega^2\rangle - \langle\Delta \omega_{\text{CSA}} ^2\rangle =465$ $(\text{krad/s})^2$, considering the total carbon Full Width at Half Maximum (FWHM) of $(5.9)^2 * 4\pi^2$ $ (\text{krad/s})^2$, the broadening due to CSA $(4.8)^2 * 4\pi^2$ $ (\text{krad/s})^2$ and neglecting carbon-carbon interactions.  Then $\langle\Delta \omega_{\perp \text{HC}} ^2\rangle = 232.5$ $ (\text{krad/s})^2$.
From the proton line one obtains $ \langle\Delta \omega_{\text{HH}} ^2\rangle= \langle\Delta \omega^2\rangle - \langle\Delta \omega_{\text{HC}} ^2\rangle = 35066$ $ (\text{krad/s})^2$ and thus $\langle\Delta \omega_{\perp \text{HH}} ^2\rangle = 23377$ $ (\text{krad/s})^2$.

The average  hyperfine field probed by the nuclei can be calculated inside a sphere centred on the radical (see Subsection 4.3).  The calculation yields the result of 8.2$\cdot 10^{-4}$ Tesla for both nuclei assuming the same electron-nucleus distance, thus for protons $\langle\Delta \omega_{\perp \text{eH}} ^2\rangle= 2/5(\mu_0/4\pi)^2\gamma_H ^2 \gamma_e ^2  \hbar^2 S(S+1) \left \langle 1/r^{6}_{eH}\right \rangle=19172$ $ (\text{krad/s})^2$ and for carbons $\langle\Delta \omega_{\perp \text{eC}} ^2\rangle= 2/5(\mu_0/4\pi)^2\gamma_C ^2 \gamma_e ^2  \hbar^2 S(S+1) \left \langle 1/r^{6}_{eC} \right \rangle=1198$ $ (\text{krad/s})^2$.

For the calculation of $1/T_{\text{1H}}$ at 0.87 Tesla $\omega_H/2\pi=37.02$ $\text{MHz}$ and $\omega_C\simeq 1/4\omega_H$. For the calculation of  $1/T_{\text{1C}}$ at 3.46 Tesla $\omega_C/2\pi=37.02$ $\text{MHz}$ and $\omega_H\simeq 4 \omega_C$. The ratio $T_{\text{1C}}(T)/T_{\text{1H}}(T)$ is then estimated taking into account the experimental condition $\omega_L=\omega_H=\omega_C$.
For PA, starting form the model of  Eq. \ref{eq2} and \ref{eq4} and  considering $\langle\Delta \omega_{\text{CH}} ^2\rangle=\langle\Delta \omega_{\text{HC}} ^2\rangle$ in the slow motion regime the ratio reduces to

\begin{equation}
\begin{split}\label{eq21}
\frac{1/T_{\text{1H}}}{1/T_{\text{1C}}}  = & 2\frac{\langle\Delta \omega_{\perp \text{HH}} ^2\rangle}{(\frac{1}{9}+3+\frac{6}{25})\langle\Delta \omega_{\perp \text{HC}} ^2\rangle+\frac{2}{15}\omega_L^2 \Delta \sigma^2}+ \\& +  \frac{(\frac{1}{9}+3+\frac{96}{25})\langle\Delta \omega_{\perp \text{HC}} ^2\rangle}{(\frac{1}{9}+3+\frac{6}{25})\langle\Delta \omega_{\perp \text{HC}} ^2\rangle+\frac{2}{15}\omega_L^2 \Delta \sigma^2} = \\ & = 54.
\end{split}
\end{equation}

The calculation retrieves a value very close to the average $T_{\text{1C}}(T)/T_{\text{1H}}(T)\simeq 53$ estimated from the experimental data for  $T<4.2$ K, confirming that $1/T_{\text{1}}$ models in Eq. \ref{eq2} for $^1$H and in Eq. \ref{eq4} for
$^{13}$C are correct.

In the doped sample PA15 it is necessary to introduce also the hyperfine contribution to both 1-$^{13}$C and $^{1}$H relaxation. On taking into account the hyperfine dipolar interactions one has

\begin{equation}
\begin{split}\label{eq21}
\frac{1/T_{\text{1H}}}{1/T_{\text{1C}}}  = & 2\frac{\langle\Delta \omega_{\perp \text{HH}} ^2\rangle}{(\frac{1}{9}+3+\frac{6}{25})\langle\Delta \omega_{\perp \text{HC}} ^2\rangle+\frac{2}{15}\omega_L^2 \Delta \sigma^2+\langle\Delta \omega_{\perp \text{eC}} ^2\rangle}+ \\& + \frac{(\frac{1}{9}+3+\frac{96}{25})\langle\Delta \omega_{\perp \text{HC}} ^2\rangle}{(\frac{1}{9}+3+\frac{6}{25})\langle\Delta \omega_{\perp \text{HC}} ^2\rangle+\frac{2}{15}\omega_L^2 \Delta \sigma^2+\langle\Delta \omega_{\perp \text{eH}} ^2\rangle}+ \\ & +\frac{\langle\Delta \omega_{\perp \text{eH}} ^2\rangle}{(\frac{1}{9}+3+\frac{6}{25})\langle\Delta \omega_{\perp \text{HC}} ^2\rangle+\frac{2}{15}\omega_L^2 \Delta \sigma^2+\langle\Delta \omega_{\perp \text{eC}} ^2\rangle} = \\ & = 32.
\end{split}
\end{equation}

The measurement of $^{1}$H has been performed in the non labelled sample uPA, so the theoretical prediction should neglect the proton-carbon interactions for $^{1}$H, definitely giving $T_{\text{1C}}(T)/T_{\text{1H}}(T)=31$.
Thus upon considering the contribution to proton and carbon relaxation associated to the lattice dynamics, one finds a  theoretical value close to $T_{\text{1C}}(T)/T_{\text{1H}}(T)\simeq 28 \pm 3$, derived from the experimental data measured below 4.2 K in PA15. In absence of other experimental evidences, this could suggest that the relaxation rates can be explained by drawing upon this mechanism alone. However this results should be better regarded as an alternative explanation based on rough estimates of the fluctuating local fields and on the naive assumption of a common $\tau_c$ for all the processes. In particular, one should be aware that the calculation of the dipolar field produced by the electrons on the nuclei does not take into account the effect of nuclear spin diffusion (see Section 4.3). On the other hand, the behaviour of $(1/T_{\text{1C}})_{el}$ as a function of T (Fig. 11) is rather suitably reproduced by a TM model which does not imply the introduction of any tunable parameter and which is able  to account for the polarization build up rates, thus providing a cleaner interpretation scheme for the electron contribution to the $^{13}C$ nuclear relaxation. Moreover, a sizeable contribution to spin-lattice relaxation due to TM must be invoked also to explain a dependence of $(1/T_{\text{1C}}(T))$ on the cooling rate sensibly weaker in PA15 than in PA (see Appendix 6.1). If in PA15 $(1/T_{\text{1C}}(T))_{el}$ contained spectral density terms only connected to the glassy dynamics, one would rather have expected the same variation of $(1/T_{\text{1C}}(T))$ with the cooling rate in both samples.

\section*{Acknowledgements}
We gratefully acknowledge A. Rigamonti for fruitful
discussions and Albeda Research for their contributions to DNP sample preparation.
This study has been supported in part by the COST Action TD1103 (European Network for Hyperpolarization Physics and Methodology in NMR and MRI) and by Regione Piemonte (Misura II.3 del Piano Straordinario per l' Occupazione). Moreover this work has been  supported by a public grant from the ''Laboratoire d'Excellence Physics Atom Light Mater'' (LabEx PALM) overseen by the French National Research Agency (ANR) as part of the ''Investissements d'Avenir'' program (reference: ANR-10-LABX-0039).

\footnotesize{
\bibliography{biblio_new} 
\bibliographystyle{rsc} 
}

\end{document}